\documentclass[cameraready]{Interspeech}

\usepackage{cite}
\usepackage{array,booktabs,tabularx,colortbl,xcolor}

\newcommand{\first}[1]{\multicolumn{1}{>{\columncolor[rgb]{1.0,0.7,0.7}}c}{#1}}
\newcommand{\second}[1]{\multicolumn{1}{>{\columncolor[rgb]{1.0,0.85,0.75}}c}{#1}}
\newcommand{\third}[1]{\multicolumn{1}{>{\columncolor[rgb]{1.0,1.0,0.8}}c}{#1}}
\newcolumntype{C}{>{\centering\arraybackslash}X}
\newcommand{\spm}[1]{{\tiny$\pm$#1}}

\title{MeanVoiceFlow2: Joint Optimization of Mean Flow and Content Encoder\\
  for Fast One-Step Zero-Shot Voice Conversion}

\author[orcid=0009-0000-8016-5144]{Takuhiro}{Kaneko}
\author[orcid=0000-0003-3102-0162]{Hirokazu}{Kameoka}
\author[orcid=0009-0003-7107-607X]{Kou}{Tanaka}
\author[orcid=0009-0009-8186-5981]{Yuto}{Kondo}

\address{NTT, Inc., Japan}

\email{takuhiro.kaneko@ntt.com}


\begin{document}

\maketitle

\begin{abstract}
  Flow-matching approaches to voice conversion (VC) have gained attention owing to their high speech quality and strong speaker similarity. Among them, one-step models such as MeanVoiceFlow are particularly attractive because they enable efficient inference; however, their reliance on a computationally intensive content encoder remains a bottleneck. We therefore propose \textit{MeanVoiceFlow2}, a framework that jointly optimizes a flow-based conversion module and a computationally efficient content encoder. The model is trained through conversion distillation using MeanVoiceFlow and the reconstruction of real data. We further incorporate diffusion-GAN training with sample mixing and teacher-guided conditioning augmentation to enhance realism and disentanglement. Experiments on zero-shot VC showed that \textit{MeanVoiceFlow2} achieved higher perceptual quality and approximately $9\times$ faster inference than MeanVoiceFlow while maintaining comparable speaker similarity.%
  \footnote{Audio samples are available at \url{https://www.kecl.ntt.co.jp/people/kaneko.takuhiro/projects/meanvoiceflow2/}.}
\end{abstract}

\section{Introduction}
\label{sec:introduction}

Voice conversion (VC) transforms the voice of a source speaker into that of a target speaker while preserving the linguistic content. 
In particular, nonparallel (zero-shot) VC approaches have attracted attention owing to their flexibility in practical scenarios. 
Although training without explicit pairwise supervision is challenging, recent advances in deep generative models (e.g.,~\cite{DKingmaICLR2014,DRezendeICML2014,IGoodfellowNIPS2014,OAaronNIPS2017,LDinhICLRW2015}) have improved performance (e.g.,~\cite{CHsuIS2017,HKameokaTASLP2019,TKanekoEUSIPCO2018,HKameokaSLT2018,KQianICML2019,YHChenICASSP2021,JSerraNeurIPS2019}). 
Recent diffusion~\cite{JSohlICML2015,YSongNeurIPS2019,JHoNeurIPS2020} and flow-matching~\cite{YLipmanICLR2023,XLiuICLR2023,MAlbergoICLR2023} models have further advanced VC~\cite{HKameokaTASLP2024,SLiuASRU2021,VPopovICLR2022,HYChoiIS2023,HYChoiAAAI2024,JYaoAAAI2025,HYChoiICASSP2025,JZuoICASSP2025,PRenIS2025,HKameokaTASLP2025}, achieving high speech quality and strong speaker similarity.
However, most approaches rely on multi-step generation, which results in slow inference.

One-step diffusion- and flow-based methods~\cite{TKanekoIS2024,TKanekoIS2025b,TKanekoICASSP2026} have been proposed to address this issue. 
These include MeanVoiceFlow~\cite{TKanekoICASSP2026}, an approach based on Mean Flow~\cite{ZGengNeurIPS2025}.
Unlike other one-step models~\cite{TKanekoIS2024,TKanekoIS2025b}, MeanVoiceFlow can be trained entirely from scratch without external pretrained modules (e.g., a pretrained neural vocoder).
However, despite the acceleration of the main flow process, it still requires a computationally intensive content encoder, which remains a primary computational bottleneck. 
In practice, the inference time of a content encoder is approximately ten times that of a single flow step.

To address this issue, we propose \textit{MeanVoiceFlow2}, which jointly optimizes a flow-based conversion module and a content encoder.
First, the model is trained through conversion distillation using MeanVoiceFlow~\cite{TKanekoICASSP2026} to align its behavior with that of the teacher, together with real-data reconstruction to improve generation fidelity.
Second, diffusion-GAN training~\cite{ZWangICLR2023} with sample mixing is used to improve the realism without relying on external modules. 
Third, teacher-guided conditioning augmentation is used to enhance content--speaker disentanglement.

FasterVoiceGrad~\cite{TKanekoIS2025} also jointly optimizes the conversion module and content encoder.
However, it relies on a pretrained neural vocoder (e.g., \cite{JKongNeurIPS2020}) for stable adversarial training, whereas our method requires no additional pretrained modules beyond the teacher model trained on the same data.
Furthermore, FasterVoiceGrad is based on a diffusion formulation, whereas our approach is based on Mean Flow.

Experiments on zero-shot (nonparallel any-to-any) VC demonstrated the effectiveness of each component and showed that \textit{MeanVoiceFlow2} achieved higher perceptual quality and significantly faster inference than MeanVoiceFlow while maintaining a comparable speaker similarity.

The remainder of this paper is organized as follows:
Section~\ref{sec:meanvoiceflow} reviews MeanVoiceFlow. 
Section~\ref{sec:meanvoiceflow2} presents \textit{MeanVoiceFlow2}. 
Section~\ref{sec:experiments} presents our experimental results. 
Finally, Section~\ref{sec:conclusion} concludes the paper.

\section{Preliminary: MeanVoiceFlow}
\label{sec:meanvoiceflow}

In this section, we briefly review MeanVoiceFlow~\cite{TKanekoICASSP2026}, which serves as the basis of \textit{MeanVoiceFlow2}. 
MeanVoiceFlow is a one-step zero-shot VC model based on Mean Flow~\cite{ZGengNeurIPS2025}.
For further details, refer to the original paper~\cite{TKanekoICASSP2026}.

MeanVoiceFlow models a mel-spectrogram $x \sim p_{\mathrm{data}}(x)$ and performs conversion conditioned on speaker embedding $s$ and content embedding $c$, which are extracted by a speaker encoder (e.g.,~\cite{YJiaNeurIPS2018}) and a content encoder (e.g.,~\cite{SLiuTASPL2021}), respectively. 
During the conversion, speaker embedding $s^{\mathrm{tgt}}$ is extracted from the target speech and content embedding $c^{\mathrm{src}}$ is extracted from the source speech. 
Here, the superscripts $\mathrm{src}$ and $\mathrm{tgt}$ denote source and target speakers, respectively.

MeanVoiceFlow considers the following linear flow path:
\begin{flalign}
  z_t^{\mathrm{tgt}} = (1 - t) x_{\mathrm{real}}^{\mathrm{tgt}} + t \hat{\epsilon}^{\mathrm{src}},
\end{flalign}
where $t \in [0,1]$ is the time. 
$\hat{\epsilon}^{\mathrm{src}}$ indicates the diffused source sample and is defined as follows:
\begin{flalign}
  \hat{\epsilon}^{\mathrm{src}} = (1 - \alpha) x_{\mathrm{real}}^{\mathrm{src}} + \alpha \epsilon,
\end{flalign}
where, $\epsilon \sim \mathcal{N}(0, I)$.
During training, $\alpha \in [0,1]$ is sampled from a logit-normal distribution~\cite{PEsserICML2024}, that is, $\alpha' \sim \mathcal{N}(0,1)$ and $\alpha = \sigma(\alpha')$, where $\sigma(\cdot)$ denotes the sigmoid function.
During inference, $\alpha$ is fixed to a constant value.
Using $\hat{\epsilon}^{\mathrm{src}}$ instead of pure noise $\epsilon$ yields a unified formulation that covers both unconditional generation and source-conditioned conversion.

Let $v(z_t, t, s, c, \alpha)$ denote the instantaneous velocity at $(z_t, t)$ conditioned on $s$, $c$, and $\alpha$. 
MeanVoiceFlow employs the average velocity over the interval $[r, t]$,
\begin{flalign}
  u(z_t, r, t, s, c, \alpha) 
  = \frac{1}{t - r} \int_r^t v(z_\tau, \tau, s, c, \alpha)\, d\tau,
\end{flalign}
which represents the average displacement between $r$ and $t$.

The average velocity is modeled by a neural network $u_{\theta}$. 
One-step conversion is performed as follows:
\begin{flalign}
  \label{eq:conversion}
  x_{\theta}^{\mathrm{conv}} 
  = \hat{\epsilon}^{\mathrm{src}} 
  - u_{\theta} \left( \hat{\epsilon}^{\mathrm{src}}, 0, 1, s^{\mathrm{tgt}}, c^{\mathrm{src}}, \alpha \right),
\end{flalign}
where $s^{\mathrm{tgt}}$ is obtained by randomly shuffling $s^{\mathrm{src}}$ within a mini-batch during training. 
For simplicity, we denote $u_{\theta} \left( \hat{\epsilon}^{\mathrm{src}}, 0, 1, s^{\mathrm{tgt}}, c^{\mathrm{src}}, \alpha \right)$ by $u_{\theta}^{\mathrm{conv}}$.

\section{Proposal: MeanVoiceFlow2}
\label{sec:meanvoiceflow2}

In MeanVoiceFlow, the content embedding $c$ is provided as an input to the model. 
However, extracting $c$ requires a computationally intensive content encoder (e.g., a Conformer~\cite{AGulatiIS2020}-based model~\cite{SLiuTASPL2021}), which is the main computational bottleneck, as discussed in Section~\ref{sec:introduction}.%
\footnote{The speaker encoder is processed only once prior to conversion and therefore does not constitute a bottleneck.}

\begin{figure}[t]
  \centering
  \includegraphics[width=\linewidth]{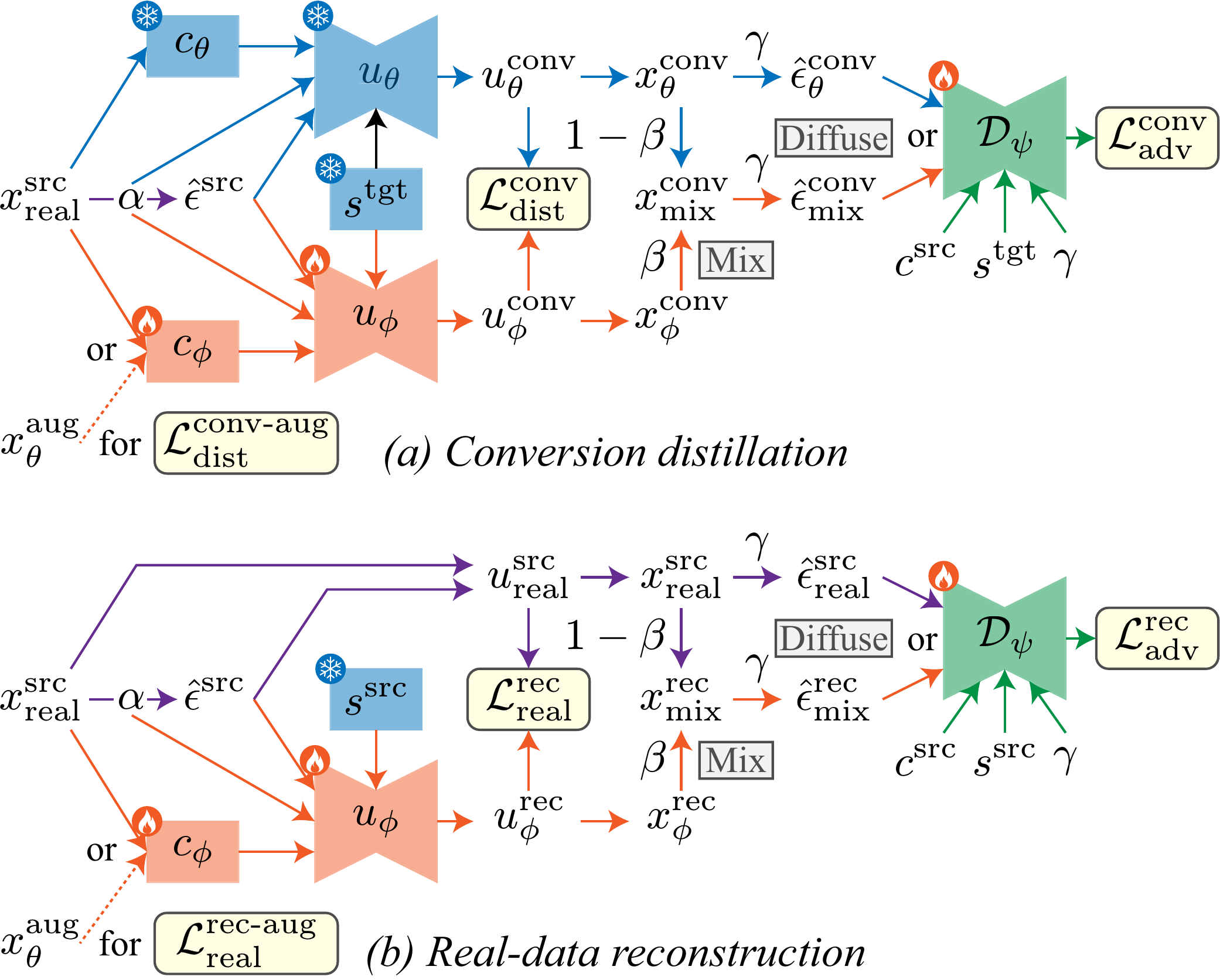}
  \vspace{-2mm}
  \caption{Overview of \textit{MeanVoiceFlow2}.
    The student jointly learns a computationally efficient content encoder $c_{\phi}$ and an average velocity network $u_{\phi}$ through (a) conversion distillation and (b) real-data reconstruction. 
    We further incorporate diffusion-GAN training with sample mixing using a discriminator $\mathcal{D}_{\psi}$ to promote realism, and teacher-guided conditioning augmentation based on $x_{\theta}^{\mathrm{aug}}$ to promote disentanglement.}
  \label{fig:overview}
\end{figure}

To address this issue, we replace the fixed pretrained content encoder, $c_{\theta}$, with a computationally efficient and trainable content encoder, $c_{\phi}$, as shown in Figure~\ref{fig:overview}. 
In addition, we introduce a student average velocity network $u_{\phi}$, which is jointly optimized with $c_{\phi}$ under the guidance of the pretrained teacher model $u_{\theta}$. 
Here, $\theta$ denotes the fixed teacher parameters, and $\phi$ represents the trainable student parameters. 
This design enables end-to-end training while eliminating the computational bottleneck caused by a fixed content encoder.

From a training perspective, \textit{MeanVoiceFlow2} comprises three key components:
(1) \textit{joint conversion distillation and real-data reconstruction},
(2) \textit{diffusion-GAN training with sample mixing}, and
(3) \textit{teacher-guided conditioning augmentation}.

\subsection{Joint conversion distillation and reconstruction}
\label{subsec:conversion_reconstruction}

\noindent\textbf{Conversion distillation.}
We introduce the following conversion distillation loss to mimic the input--output behavior of the teacher model:
\begin{flalign}
  \mathcal{L}_{\mathrm{dist}}^{\mathrm{conv}}(\phi)
  = \mathbb{E} \left[
  d \left( u_{\theta}^{\mathrm{conv}}, u_{\phi}^{\mathrm{conv}} \right)
  \right].
\end{flalign}
Here, $u_{\theta}^{\mathrm{conv}}$ and $u_{\phi}^{\mathrm{conv}}$ denote the teacher and student outputs, respectively, in the conversion setting, where the teacher output is defined using Eq.~\ref{eq:conversion}. 
The student output is defined as follows:
\begin{flalign}
  \label{eq:u_phi_conv}
  u_{\phi}^{\mathrm{conv}} 
  = u_{\phi} \left( \hat{\epsilon}^{\mathrm{src}}, 0, 1, s^{\mathrm{tgt}}, c_{\phi}(x_{\mathrm{real}}^{\mathrm{src}}), \alpha \right).
\end{flalign}
The function $d(\cdot,\cdot)$ denotes the distance measure. 
Following prior work~\cite{ZGengNeurIPS2025,TKanekoICASSP2026}, we adopt an adaptively weighted loss~\cite{ZGengICLR2025},
$d(a, b) = \frac{\lVert a - b \rVert_2^2} {\mathrm{sg} \left( \lVert a - b \rVert_2^2 + \varepsilon \right)}$,
where $\mathrm{sg}(\cdot)$ denotes the stop-gradient operation and $\varepsilon = 10^{-3}$ is used to prevent division by zero.

\smallskip
\noindent\textbf{Real-data reconstruction.}
In the distillation framework above, the student is supervised by the teacher-generated conversion output.
However, distillation alone does not directly ensure consistency with real data.
To address this without relying on paired source--target data in a nonparallel setting, we introduce the following real-data reconstruction loss:
\begin{flalign}
  \mathcal{L}_{\mathrm{real}}^{\mathrm{rec}}(\phi)
  = \mathbb{E} \left[
  d \left( u_{\phi}^{\mathrm{rec}}, u_{\mathrm{real}}^{\mathrm{src}} \right)
  \right].
\end{flalign}
Here, $u_{\phi}^{\mathrm{rec}}$ denotes the student output under the reconstruction condition, and is defined as follows:
\begin{flalign}
  \label{eq:u_phi_rec}
  u_{\phi}^{\mathrm{rec}}
  = u_{\phi} \left( \hat{\epsilon}^{\mathrm{src}}, 0, 1, s^{\mathrm{src}}, c_{\phi}(x_{\mathrm{real}}^{\mathrm{src}}), \alpha \right).
\end{flalign}
The ground-truth velocity derived from real source data is
\begin{flalign}
  u_{\mathrm{real}}^{\mathrm{src}}
  = \hat{\epsilon}^{\mathrm{src}} - x_{\mathrm{real}}^{\mathrm{src}}.
\end{flalign}

\subsection{Diffusion-GAN training with sample mixing}
\label{subsec:diffusion_gan_mixing}

\noindent\textbf{Adversarial conversion training.}
Using only the pairwise distance loss often leads to statistical averaging, which degrades perceptual realism.
To address this issue, we introduce adversarial training based on a diffusion-GAN~\cite{ZWangICLR2023}, which promotes distribution-level alignment while stabilizing optimization.
Additionally, we incorporate data mixing to smooth the discriminator decision boundary and emphasize the subtle discrepancies between the teacher and student samples.

Eq.~\ref{eq:conversion} is used to obtain teacher- and student-generated samples $x_{\theta}^{\mathrm{conv}}$ and $x_{\phi}^{\mathrm{conv}}$, respectively.
The teacher samples are regarded as real examples to guide the student distribution toward the teacher distribution.
We then construct mixed samples as follows:
\begin{equation}
  x_{\mathrm{mix}}^{\mathrm{conv}}
  = (1 - \beta) x_{\theta}^{\mathrm{conv}}
  + \beta x_{\phi}^{\mathrm{conv}},
\end{equation}
where $\beta \in [0,1]$ is drawn from a logit-normal distribution~\cite{PEsserICML2024} in the same manner as $\alpha$.

Both the teacher and mixed samples are further perturbed by the diffusion process~\cite{ZWangICLR2023}:
\begin{align}
  \hat{\epsilon}_{\theta}^{\mathrm{conv}} 
  &= (1 - \gamma) x_{\theta}^{\mathrm{conv}} + \gamma \epsilon^{\prime}, \\
  \hat{\epsilon}_{\mathrm{mix}}^{\mathrm{conv}} 
  &= (1 - \gamma) x_{\mathrm{mix}}^{\mathrm{conv}} + \gamma \epsilon^{\prime},
\end{align}
where $\epsilon^{\prime} \sim \mathcal{N}(0, I)$ and $\gamma \in [0,1]$ is sampled from a logit-normal distribution.

We adopt a least-squares GAN objective~\cite{XMaoICCV2017},
\begin{flalign}
  \mathcal{L}_{\mathrm{adv}}^{\mathrm{conv}}(\psi) 
  &= \mathbb{E} \Big[
    (\mathcal{D}_{\psi}(\hat{\epsilon}_{\theta}^{\mathrm{conv}}, s^{\mathrm{tgt}}, c^{\mathrm{src}}, \gamma) - 1)^2 \nonumber \\
  &\quad + (\mathcal{D}_{\psi}(\hat{\epsilon}_{\mathrm{mix}}^{\mathrm{conv}}, s^{\mathrm{tgt}}, c^{\mathrm{src}}, \gamma))^2
    \Big], \\
  \mathcal{L}_{\mathrm{adv}}^{\mathrm{conv}}(\phi) 
  &= \mathbb{E} \Big[
    (\mathcal{D}_{\psi}(\hat{\epsilon}_{\mathrm{mix}}^{\mathrm{conv}}, s^{\mathrm{tgt}}, c^{\mathrm{src}}, \gamma) - 1)^2
    \Big],
\end{flalign}
where $\mathcal{D}_{\psi}$ denotes the discriminator.
This GAN objective is conditioned on $s^{\mathrm{tgt}}$, $c^{\mathrm{src}}$, and $\gamma$ to maintain consistency with the conversion setting and diffusion noise level.
In practice, the discriminator $\mathcal{D}_{\psi}$ adopts the same network architecture as the teacher average velocity network $u_{\theta}$ with time variables $r$ and $t$ omitted.

\smallskip
\noindent\textbf{Adversarial reconstruction training.}
We apply the same diffusion-GAN procedure with the data-mixing strategy to the reconstructed samples, $x_{\phi}^{\mathrm{rec}} = \hat{\epsilon}^{\mathrm{src}} - u_{\phi}^{\mathrm{rec}}$, and corresponding real data, $x_{\mathrm{real}}^{\mathrm{src}}$. 
The mixing and diffusion procedures are illustrated in Figure~\ref{fig:overview}(b).
The resulting adversarial losses are denoted as $\mathcal{L}_{\mathrm{adv}}^{\mathrm{rec}}(\psi)$ and $\mathcal{L}_{\mathrm{adv}}^{\mathrm{rec}}(\phi)$.

\subsection{Teacher-guided conditioning augmentation}
\label{subsec:condition_augmentation}

VC aims to modify speaker identity while preserving linguistic content, which requires effective content--speaker disentanglement.
To encourage the content encoder to learn speaker-invariant representations, we introduce a teacher-guided conditioning augmentation strategy.

The key idea is that the content encoder should produce consistent representations even when speaker identity changes but the underlying linguistic content remains unchanged.
To promote this property, we introduce an additional forward path in which the input to the content encoder is replaced with the speech generated by the teacher model under speaker-augmented conditions.

Specifically, we generate augmented samples using the teacher model:
\begin{equation}
  x_{\theta}^{\mathrm{aug}} 
  = \epsilon^{\prime\prime} 
  - u_{\theta}(\epsilon^{\prime\prime}, 0, 1, s^{\mathrm{aug}}, c^{\mathrm{src}}, 1),
\end{equation}
where $\epsilon^{\prime\prime} \sim \mathcal{N}(0, I)$ and $s^{\mathrm{aug}}$ is obtained by randomly shuffling $s^{\mathrm{src}}$ within a mini-batch.

The resulting representation $c_{\phi}(x_{\theta}^{\mathrm{aug}})$ replaces $c_{\phi}(x_{\mathrm{real}}^{\mathrm{src}})$ in Eqs.~\ref{eq:u_phi_conv} and~\ref{eq:u_phi_rec} as shown in Figure~\ref{fig:overview}.
Conversion distillation and real-data reconstruction are performed in the same manner as described above.
The corresponding losses are denoted as $\mathcal{L}_{\mathrm{dist}}^{\mathrm{conv\text{-}aug}}(\phi)$ and $\mathcal{L}_{\mathrm{real}}^{\mathrm{rec\text{-}aug}}(\phi)$, respectively.

\smallskip
\noindent\textbf{Final objective.}
The final student objective is as follows:
\begin{align}
  \label{eq:loss_mvf2}
  \mathcal{L}_{\mathrm{MVF2}}(\phi)
  &= \underbrace{
    \mathcal{L}_{\mathrm{dist}}^{\mathrm{conv}}(\phi)
    + \mathcal{L}_{\mathrm{dist}}^{\mathrm{conv\text{-}aug}}(\phi)
    + \lambda_{\mathrm{adv}} \mathcal{L}_{\mathrm{adv}}^{\mathrm{conv}}(\phi)
    }_{\text{Conversion}} \nonumber \\
  &+
    \underbrace{
    \mathcal{L}_{\mathrm{real}}^{\mathrm{rec}}(\phi)
    + \mathcal{L}_{\mathrm{real}}^{\mathrm{rec\text{-}aug}}(\phi)
    + \lambda_{\mathrm{adv}} \mathcal{L}_{\mathrm{adv}}^{\mathrm{rec}}(\phi)
    }_{\text{Reconstruction}},
\end{align}
where $\lambda_{\mathrm{adv}}$ is a weighting hyperparameter, which had a fixed value of $1$ in all of the experiments.
The corresponding discriminator objective is expressed as follows:
\begin{align}
  \mathcal{L}_{\mathrm{MVF2}}(\psi)
  = \mathcal{L}_{\mathrm{adv}}^{\mathrm{conv}}(\psi)
  + \mathcal{L}_{\mathrm{adv}}^{\mathrm{rec}}(\psi).
\end{align}

\section{Experiments}
\label{sec:experiments}

\subsection{Experimental setup}
\label{subsec:experimental_setup}

\noindent\textbf{Data.}
We evaluated \textit{MeanVoiceFlow2} on zero-shot (nonparallel any-to-any) VC tasks.
The experimental protocol followed that of MeanVoiceFlow~\cite{TKanekoICASSP2026}, which served as the teacher model.
Our primary experiments were conducted using the VCTK dataset~\cite{JYamagishiVCTK2019}, which contains recordings from 110 English speakers.
To examine whether our findings are robust across datasets, we also conducted experiments using LibriTTS~\cite{HZenIS2019}, which contains recordings from 1,151 English speakers.
To simulate unseen-to-unseen conversion, we held out ten speakers and ten sentences from the training for evaluation.
All audio clips were downsampled to 22.05\,kHz.
We extracted 80-dimensional log-mel spectrograms using an FFT size of 1,024, hop size of 256, and window size of 1,024, which served as the conversion targets.

\smallskip
\noindent\textbf{Implementation.}
To isolate the effects of the proposed training strategy, we adopted the same architectures as that used in previous studies~\cite{TKanekoIS2025,TKanekoICASSP2026}.
The velocity networks $u_{\theta}$ and $u_{\phi}$ were implemented as a U-Net~\cite{ORonnebergerMICCAI2015} with 12 convolutional layers (512 channels), two down/up-sampling stages, gated linear units (GLUs)~\cite{YDauphinICML2017}, and weight normalization (WN)~\cite{TSalimansNIPS2016}.
The discriminator $\mathcal{D}_{\psi}$ shared the same architecture with the time variables omitted.
The proposed content encoder $c_{\phi}$ comprises three convolutional layers (512 channels), GLUs, instance normalization~\cite{DUlyanovArXiv2016}, and WN.
For the teacher model, content embeddings were extracted using a pretrained bottleneck feature extractor~\cite{SLiuTASPL2021}, and speaker embeddings were obtained from a pretrained speaker encoder~\cite{YJiaNeurIPS2018}.
The waveforms were synthesized using HiFi-GAN V1~\cite{JKongNeurIPS2020}.
The teacher and student models were trained using Adam~\cite{DPKingmaICLR2015} (batch size $32$, learning rate $2\times10^{-4}$, $\beta_1=0.5$, and $\beta_2=0.9$).
The teacher was trained for 500 epochs, and the student and discriminator were trained for 250 epochs, initialized by the pretrained teacher.
At the inference time, only $c_{\phi}$ and $u_{\phi}$ were used, without the pretrained bottleneck extractor, which eliminated the computational overhead of the heavy fixed content encoder required by the teacher.

\smallskip
\noindent\textbf{Evaluation metrics.}
The conversion performance was evaluated using five objective metrics.
The perceptual quality was measured using three predicted mean opinion score (MOS) metrics:
\texttt{UT}$\uparrow$ (UTMOS~\cite{TSaekiIS2022}) and \texttt{DNSP}$\uparrow$ (DNSMOS Pro~\cite{FCumlinIS2024}, trained on BVCC~\cite{WHHuangIS2022}) for synthesized/converted speech, and \texttt{DNS}$\uparrow$ (DNSMOS~\cite{CReddyICASSP2021}) for noise-suppressed speech.
We also used \texttt{CER}$\downarrow$ with Whisper-large-v3~\cite{ARadfordICML2023} for intelligibility and \texttt{SECS}$\uparrow$ with WavLM Base+~\cite{SChenJSTSP2022} for speaker similarity.
All of the metrics were computed for 8,100 speaker--sentence pairs.
For the key comparisons in Section~\ref{subsec:comparison_previous}, we conducted additional subjective evaluations and speed assessments.

\subsection{Component analysis}
\label{subsec:component_analysis}

\noindent\textbf{Analysis of joint conversion and reconstruction.}
We analyzed the effects of joint conversion and reconstruction.
Table~\ref{tab:analysis_conversion_reconstruction} summarizes the ablations in which the conversion- and reconstruction-related components were removed.
Conditioning augmentation was not performed to isolate its effects.
(i) Full model (c) achieved the best overall performance across most metrics.
(ii) Removing reconstruction (a) degraded perceptual quality (\texttt{UT}, \texttt{DNSP}) and intelligibility (\texttt{CER}), indicating that reconstruction helped align the generated distribution with the real-data distribution beyond teacher guidance.
(iii) Removing conversion (b) yielded a low \texttt{CER} but severely degraded speaker similarity (\texttt{SECS}) and perceptual quality (\texttt{UT}, \texttt{DNSP}), which suggests that the model largely preserved the input speech characteristics rather than performing proper VC.

\begin{table}[h]
  \caption{Analysis of joint conversion and reconstruction.
    Conv and Rec indicate the use of conversion distillation and real-data reconstruction, respectively.}
  \vspace{-3mm}
  \label{tab:analysis_conversion_reconstruction}
  \setlength{\tabcolsep}{2pt}
  \centering
  \scriptsize{
    \begin{tabularx}{\linewidth}{cccCCCCC}
      \toprule
      & Conv & Rec &
      \texttt{UT$\uparrow$} & \texttt{DNSP$\uparrow$} & \texttt{DNS$\uparrow$} & \texttt{CER$\downarrow$} & \texttt{SECS$\uparrow$}
      \\ \midrule
      (a) & $\checkmark$ & &
      \second{4.00} & \second{2.94} & \first{3.80} & \third{1.8} & \first{0.885}
      \\
      (b) & & $\checkmark$  &
      \third{3.60} & \third{2.40} & \third{3.77} & \first{0.1} & \third{0.641}
      \\
      (c) & $\checkmark$ & $\checkmark$ &
      \first{4.04} & \first{2.97} & \first{3.80} & \second{1.5} & \first{0.885}
      \\ \bottomrule
    \end{tabularx}
  }
  \vspace{-4mm}
\end{table}

\smallskip
\noindent\textbf{Analysis of adversarial training.}
The proposed adversarial framework comprises
(1) diffusion-GAN training,
(2) sample mixing, and
(3) no reliance on external modules (e.g., a pretrained neural vocoder), unlike FastVoiceGrad~\cite{TKanekoIS2024,TKanekoIS2025b}, which employs a pretrained vocoder with either a waveform discriminator (WD)~\cite{SLeeICLR2023} or a vocoder-projected feature discriminator (VPFD)~\cite{TKanekoIS2025b}.
Ablation was performed to evaluate each component.
Conditioning augmentation was not performed to isolate its effects.
Table~\ref{tab:analysis_adversarial_training} summarizes these results.
(i) A comparison between (a) and (e) shows that adversarial training improved perceptual quality (\texttt{DNSP}, \texttt{DNS}), intelligibility (\texttt{CER}), and speaker similarity (\texttt{SECS}) over the non-adversarial baseline while maintaining \texttt{UT}.
(ii) A comparison between (b) and (c) shows that diffusion improved \texttt{UT} and reduced \texttt{CER}, highlighting its role in preserving speech quality.
(iii) A comparison between (b) and (d) shows that sample mixing improved \texttt{DNSP} and \texttt{DNS}, suggesting a regularization effect.
(iv) Full configuration (e) achieved the best overall.
(v) Compared with WD (f) and VPFD (g), the proposed method (e) achieved competitive or superior performance without a pretrained neural vocoder.

\begin{table}[h]
  \vspace{-2mm}
  \caption{Analysis of adversarial training.
    Diffuse and Mix indicate the use of diffusion-GAN training and sample mixing, respectively.}
  \vspace{-3mm}
  \label{tab:analysis_adversarial_training}
  \setlength{\tabcolsep}{2pt}
  \centering
  \scriptsize{
    \begin{tabularx}{\linewidth}{ccccCCCCC}
      \toprule
      & GAN & Diffuse & Mix &
      \texttt{UT$\uparrow$} & \texttt{DNSP$\uparrow$} & \texttt{DNS$\uparrow$} & \texttt{CER$\downarrow$} & \texttt{SECS$\uparrow$}
      \\ \midrule
      (a) & None & -- & -- &
      \first{4.04} & 2.79 & 3.75 & 2.2 & 0.882
      \\ \midrule
      (b) & Proposed & & &
      4.01 & 2.87 & \third{3.79} & 1.9 & \third{0.883}
      \\
      (c) & Proposed & $\checkmark$ & &
      \first{4.04} & 2.86 & 3.78 & \first{1.5} & 0.882
      \\
      (d) & Proposed & & $\checkmark$ &
      3.93 & \second{2.90} & \first{3.80} & 1.9 & 0.882
      \\
      (e) & Proposed& $\checkmark$ & $\checkmark$ &
      \first{4.04} & \first{2.97} & \first{3.80} & \first{1.5} & \first{0.885}
      \\ \midrule
      (f) & WD~\cite{SLeeICLR2023} & -- & -- &
      \first{4.04} & 2.89 & \third{3.79} & \third{1.8} & \second{0.884}
      \\
      (g) & VPFD~\cite{TKanekoIS2025b} & -- & -- &
      4.02 & \second{2.90} & \third{3.79} & \third{1.8} & \second{0.884}
      \\ \bottomrule
    \end{tabularx}
  }
  \vspace{-2mm}
\end{table}

\smallskip
\noindent\textbf{Analysis of conditioning augmentation.}
We evaluated the effectiveness of the conditioning augmentation (\textit{CondAug}).
In addition to the standard ablation, we evaluated an alternative approach, Direct Distill, which adds an $\ell_1$ loss to the objective without CondAug to explicitly align the student content representation $c_{\phi}$ with the teacher representation $c_{\theta}$.
Table~\ref{tab:analysis_conditioning_augmentation} presents the results.
CondAug (b) improved all objective metrics over the w/o CondAug configuration (a).
By contrast, Direct Distill (c) did not provide comparable improvements and instead degraded \texttt{DNSP} and \texttt{CER}.
Although direct feature alignment enforced similarity to the teacher representation, it appeared to overly constrain the student content encoder.
These findings suggest that CondAug offers more effective implicit regularization than explicit feature-level distillation.

\begin{table}[h]
  \vspace{-2mm}
  \caption{Analysis of conditioning augmentation (CondAug).
    Direct Distill adds an explicit $\ell_1$ loss between the student and teacher content representations to the w/o CondAug objective.}
  \vspace{-3mm}
  \label{tab:analysis_conditioning_augmentation}
  \setlength{\tabcolsep}{2pt}
  \centering
  \scriptsize{
    \begin{tabularx}{\linewidth}{ccCCCCC}
      \toprule
      & &
      \texttt{UT$\uparrow$} & \texttt{DNSP$\uparrow$} & \texttt{DNS$\uparrow$} & \texttt{CER$\downarrow$} & \texttt{SECS$\uparrow$}
      \\ \midrule
      (a) & w/o CondAug &
      \third{4.04} & \second{2.97} & \second{3.80} & \second{1.5} & \second{0.885}
      \\
      (b) & w/ CondAug &
      \first{4.05} & \first{2.99} & \first{3.81} & \first{1.2} & \first{0.887}
      \\ \midrule
      (c) & Direct Distill &
      \first{4.05} & \third{2.94} & \second{3.80} & \third{1.9} & \third{0.884}
      \\ \bottomrule
    \end{tabularx}
  }
  \vspace{-3mm}
\end{table}

\subsection{Comparison with previous models}
\label{subsec:comparison_previous}

We compared \textit{MeanVoiceFlow2} (\textit{MVF2}) with previous models to assess its relative performance.
Specifically, we compared it with MeanVoiceFlow (MVF; the teacher model) and FasterVoiceGrad (FVG2)~\cite{TKanekoIS2025}, which also jointly distills the conversion module and content encoder but adopts a diffusion-model backbone~\cite{HKameokaTASLP2024} and relies on a pretrained neural vocoder for stable adversarial training.
We included ground-truth (GT) speech and DiffVC (30 iterations)~\cite{VPopovICLR2022} as anchor samples.
For a comprehensive evaluation, we conducted MOS tests on 90 speaker--sentence pairs per model.
Naturalness was rated on a five-point scale 
(\texttt{nMOS}: 1 = bad, 2 = poor, 3 = fair, 4 = good, and 5 = excellent).
Speaker similarity was rated on a four-point scale 
(\texttt{sMOS}: 1 = different (sure), 2 = different (not sure), 3 = same (not sure), and 4 = same (sure)).
The tests were conducted online with 12 participants, with over 1,500 ratings collected for each test.
In addition, we measured the inference speed of the main conversion flow and content encoder using a real-time factor (\texttt{RTF}) on a single NVIDIA GeForce RTX 4090 GPU.
Table~\ref{tab:comparison_previous} summarizes the results.
\textit{MVF2} achieved a higher perceptual quality (\texttt{nMOS}, \texttt{UT}, \texttt{DNSP}, \texttt{DNS}) than MVF, while maintaining comparable intelligibility (\texttt{CER}) and speaker similarity (\texttt{sMOS}, \texttt{SECS}), and significantly reducing the real-time factor (\texttt{RTF}) by approximately $9 \times$.  
Compared to FVG2, \textit{MVF2} achieved better or comparable performance across all metrics, with significant improvements in \texttt{nMOS} and \texttt{DNSP} and without relying on a pretrained neural vocoder for training.

\begin{table}[h]
  \vspace{-1mm}
  \caption{Comparison with previous models in terms of subjective metrics (nMOS and sMOS with 95\% confidence intervals), objective metrics, and RTF.  
$^*$ indicates a statistically significant difference from \textit{MVF2} on the Mann--Whitney U test ($p < 0.05$).}
  \vspace{-3mm}
  \label{tab:comparison_previous}
  \setlength{\tabcolsep}{1pt}
  \centering
  {\fontsize{6.5pt}{7pt}\selectfont
    \begin{tabularx}{\linewidth}{ccccCCCCCl}
      \toprule
      & &
      \texttt{nMOS$\uparrow$} & \texttt{sMOS$\uparrow$} & \texttt{UT$\uparrow$} & \texttt{DNSP$\uparrow$} & \texttt{DNS$\uparrow$} & \texttt{CER$\downarrow$} & \texttt{SECS$\uparrow$} & \multicolumn{1}{c}{\texttt{RTF$\downarrow$}}
      \\ \midrule
      (a) & GT &
      4.26\spm{.09}$^*$ & 3.64\spm{.06}$^*$ & 4.15 & 2.89 & 3.75 & 0.1 & 0.940 & \multicolumn{1}{c}{--}
      \\ \midrule
      (b) & DiffVC &
      3.43\spm{.11}$^*$ & 2.24\spm{.10}$^*$ & 3.76 & 2.64 & 3.75 & 5.4 & 0.880 & 0.19
      \\ \midrule
      (c) & MVF &
      \second{3.76\spm{.09}$^*$} & \first{2.74\spm{.11}\thickspace\thinspace\thinspace} & \third{3.98} & \second{2.85} & \third{3.78} & \first{1.2} & \third{0.886} & \third{0.0072~~~}
      \\
      (d) & \textit{MVF2} &
      \first{3.93\spm{.10}\thickspace\thinspace\thinspace} & \second{2.70\spm{.11}\thickspace\thinspace\thinspace} & \first{4.05} & \first{2.99} & \second{3.81} & \first{1.2} & \second{0.887} & \first{0.00084}
      \\ \midrule
      (e) & FVG2 &
      \third{3.72\spm{.10}$^*$} & \third{2.63\spm{.10}\thickspace\thinspace\thinspace} & \second{4.03} & \third{2.79} & \first{3.82} & \first{1.2} & \first{0.890} & \first{0.00084}
      \\ \bottomrule
    \end{tabularx}
  }
  \vspace{-3mm}
\end{table}

\subsection{Evaluation on LibriTTS}
\label{subsec:evaluation_libritts}

To assess whether our findings are robust across datasets, we evaluated MVF (teacher) and \textit{MVF2} (student) on LibriTTS~\cite{HZenIS2019}.
As listed in Table~\ref{tab:comparison_libritts}, \textit{MVF2} achieved better or comparable performance across all metrics, with significant improvements in \texttt{UT} and \texttt{DNSP} while reducing the real-time factor (\texttt{RTF}) by approximately $9\times$ compared to MVF.
These results demonstrate that the proposed distillation framework was effective and computationally efficient for various datasets.

\begin{table}[h]
  \vspace{-1mm}
  \caption{Comparison on LibriTTS.}
  \vspace{-3mm}
  \label{tab:comparison_libritts}
  \setlength{\tabcolsep}{1pt}
  \centering
  \scriptsize{
    \begin{tabularx}{\linewidth}{ccCCCCCC}
      \toprule
      & & \texttt{UT$\uparrow$} & \texttt{DNSP$\uparrow$} & \texttt{DNS$\uparrow$} & \texttt{CER$\downarrow$} & \texttt{SECS$\uparrow$} & \texttt{RTF$\downarrow$}
      \\ \midrule
      (a) & MVF &
      \third{3.93} & \third{3.01} & \third{3.70} & \first{1.1} & \third{0.879} & \third{0.0089}
      \\
      (b) & \textit{MVF2} &
      \first{4.05} & \first{3.10} & \first{3.71} & \first{1.1} & \first{0.880} & \first{0.0010}
      \\ \bottomrule
    \end{tabularx}
  }
  \vspace{-3.5mm}
\end{table}

\section{Conclusion}
\label{sec:conclusion}

We presented \textit{MeanVoiceFlow2}, a one-step zero-shot VC model that jointly optimizes a flow-based conversion module and an efficient content encoder.
This framework integrates conversion distillation with reconstruction, diffusion-GAN training with sample mixing, and conditioning augmentation.
Experiments showed that \textit{MeanVoiceFlow2} achieved higher perceptual quality than MeanVoiceFlow while maintaining comparable speaker similarity and reducing the inference time by approximately $9\times$.
Future work will include extending the framework to practical applications such as accent and real-time VC.

\section{Generative AI Use Disclosure}

The authors used generative AI tools to edit and polish the language of the manuscript.
The authors have verified the final manuscript and are responsible for its content.

\bibliographystyle{IEEEtran}
\bibliography{refs}

\end{document}